\documentstyle[12pt]{article}

\begin{document}
\begin{flushright}
5 December, 1997

OU-HET 285

hep-th/9712054
\end{flushright}
\vskip1.5cm
\begin{center}
{\Large {\bf 
Renormalization Group Flow 
near  the Superconformal Points
in $N=2$ Supersymmetric  Gauge Theories}}

\vskip2.5cm
{\large  Takahiro Kubota 
\footnote {e-mail: kubota@het.phys.wani.osaka-u.ac.jp} and Naoto 
Yokoi \footnote {e-mail: yokoi@het.phys.wani.osaka-u.ac.jp}}

\vskip1cm
{\it  Department of Physics, Osaka University}

{\it Toyonaka,  Osaka, 560 Japan}
\end{center}
\vskip2.5cm
\begin{center}
{\bf Abstract}
\end{center}
The behavior of the beta-function of the low-energy effective 
coupling  in  the $N=2$ supersymmetric 
$SU(2)$ QCD with several massive matter hypermultiplets and 
in the $SU(3)$ Yang-Mills theory is 
determined near the superconformal 
points in the moduli space. The renormalization group flow is 
unambiguously fixed by looking at limited types of deformation 
near the superconformal points.  
It is pointed out that the scaling dimension of the beta-function 
is controlled by the scaling behavior of moduli parameters 
and the relation between them is explicitly worked out.
Our scaling dimensions of the beta-functions are consistent 
in part with the results obtained recently by Bilal and Ferrari 
in a different method for the $SU(2)$ QCD.

\vfill\eject
\begin{flushleft}  
{\bf 1. Introduction}
\end{flushleft}
\vskip0.5cm
Among many developments that have taken place after the work of 
Seiberg and Witten [1] on the $N=2$ supersymmetric Yang-Mills 
and QCD theories (see Ref. [2] for reviews), one of the 
most intriguing  is the 
existence of the superconformal symmetry. This symmetry 
is expected to be realized at some particular points in the moduli 
space (superconformal points)
where massless particles with mutually non-local charge  exist. 
This has been observed 
in $SU(3)$ gauge group without matter hypermultiplet [3], and in 
$SU(2)$ with massive hypermultiplets [4] and variations thereof. 
There have been a lot of strong evidences  for the  existence of
such interacting non-trivial  superconformal field theories, in 
particular on the basis of 
the representation theory of the superconformal algebra. 
The classification of the superconformal points in the general 
$SU(N_{c})$ gauge group has also been done  [5].
 
In  conventional field theory techniques, the most 
straightforward way of examining the conformal invariance is to 
study the behavior of the beta-function or the renormalization 
group flow.  Although there have been several attempts [6-8] to 
derive the beta-function in the Seiberg-Witten theories, the 
idea of the renormalization group flow is yet to be investigated 
in more details.  In the present paper, we will 
inspect the behavior of the beta-function near the 
superconformal points by fitting the 
renormalization group idea {\` a} la Wilson in the Seiberg-Witten 
theories. 

In the Wilson type renormalization group, we start with an 
ultra-violet cut-off scale,  say, $M_{0}$  and with another 
intermediate energy scale $M$. We then integrate out all dynamical 
degrees of freedom in the momentum space between $M_{0}$ and $M$. 
The effective Lagrangian contains only a few relevant interaction 
terms. In the case of Seiberg-Witten theory with the $SU(2)$ gauge 
symmetry broken to $U(1)$, the low-energy effective theory is 
described by the $N=2$,  $U(1)$ gauge field multiplet 
($W_{\alpha }$, $A$);
\begin{eqnarray}
   {\cal L}&=&{\rm Im}\frac{1}{4\pi}
   \left [ \int d^4 \theta
   \frac{\partial {\cal F} (A)}{\partial A} A^{\dag}
   +\frac{1}{2}\int d^2 \theta  \tau (A) 
   W_{\alpha}W^{\alpha}\right ].
\end{eqnarray}
The function $\tau (A)$ is expressed as the second derivative 
of the prepotential, 
$\tau (A)=\partial ^{2}{\cal F}(A)/\partial A^{2}$.
 The bosonic part of $\tau (A)$ is related to  the gauge 
coupling $g^2$ and the vacuum angle $\theta $ via
\begin{eqnarray}
\tau (a)=i\frac{8\pi}{g^2}+\frac{\theta}{\pi},
\end{eqnarray}
where $a$ is the vacuum expectation value of $\phi $, $\phi $ 
being the scalar component of the gauge multiplet. If we include 
 $N_{f}$ matter hypermultiplets with bare masses $m_{i}$, ($i=1, 
\cdots , N_{f}$),  $\tau $ becomes  a function of 
$u/\Lambda ^2$ and $m_{i}/\Lambda $. Here 
$\Lambda $ is the QCD dynamical mass scale and $u$ is the expectation 
value ${\rm Tr}(\phi ^2)$.  Seiberg and Witten have determined 
$\tau (u/\Lambda ^2, m_{i}/\Lambda )$ completely on the basis of the 
elliptic curves.

We are now interested in the linear response under the change of the 
energy scale $M$. The moduli parameters $u$ and $m_{i}$ are in 
principle dependent on this scale $M$ on the non-perturbative level. 
The low-energy Seiberg-Witten action does not seem to be telling us 
anything about the procedure of integrating out the dynamical degrees 
of freedom between $M_{0}$ and $M$. Near the superconformal points, 
however,  we can still think of consistent renormalization group flows 
and discuss various critical exponents.
Argyres et al. [4] in fact considered consistent deformations  
 near the superconformal points in the $SU(2)$ QCD case and 
have determined the scaling dimensions of $u$ and $m_{i}$.
In their recent interesting papers [9], Bilal and Ferrari 
pointed out that the behavior of the beta-function near the 
superconformal points is constrained by the eigenvalues of monodromy 
matrices. They found a relation between the scaling dimensions of the 
beta-function  and $u$. 
In this paper we will study the behavior of the beta-function 
near the superconformal points in a more direct method. Our 
approach goes along the line of Argyres et al. [4] and is more 
straightforward than that of Bilal and Ferrari [9].

In some of the literatures we sometimes find a conjecture that 
the exact relation found in [10] between the moduli parameter 
$u$ and the derivative of the prepotential ${\cal F}$ 
with respect to the QCD dynamical scale $\Lambda $, i.e.,
$\Lambda (\partial {\cal F}/\partial \Lambda)$ 
might have some relevance to the renormalization group idea. In 
fact in the weak coupling region, the $\Lambda$-derivative of the 
coupling $\tau  (u/\Lambda ^{2}, m_{i}/\Lambda)$  
coincides with the coefficient of the beta-function.
As we will see shortly, however, in the strong coupling region, 
there is no guarantee that the flow under the change of the 
 energy scale $M$ and the $\Lambda$-derivative of 
$\tau (u/\Lambda ^{2}, m_{i}/\Lambda )$, or ${\cal F}$ are 
directly connected.
\vskip1cm
\begin{flushleft}
{\bf  2. Renormalization Group Flow}
\end{flushleft}
\vskip0.5cm
Let us begin with the elliptic curves in the $SU(2)$ gauge group 
with several matter hypermultiplets ($N_{f}=1,2,3$). 
They are given the following structures [1];
\begin{eqnarray}
y^2=x^2(x-u)+P_{N_{f}}(x,u,m_{i},\Lambda),
\end{eqnarray}
where
\begin{eqnarray}
P_{1}&=&\frac{1}{4}\Lambda ^3 m x-\frac{1}{64}\Lambda ^6 ,
\\
P_{2}&=&-\frac{1}{64}\Lambda ^4 (x-u)+\frac{1}{4}\Lambda ^2 
m_{1} m_{2} x - \frac{1}{64}
\Lambda ^4 (m_{1}^2+m_{2}^2),
\\
P_{3}&=&-\frac{1}{64}\Lambda ^2 (x-u)^2 - \frac{1}{64}\Lambda ^2 (
x-u)(m_{1}^2+m_{2}^2+m_{3}^2)
\nonumber \\
& &+\frac{1}{4}\Lambda m_{1}m_{2}m_{3}x -\frac{1}{64}\Lambda ^2 (
m_{1}^2 m_{2}^2 +m_{1}^2 m_{3}^2 + m_{2}^2 m_{3}^2).
\end{eqnarray}
We will use the same notation $\Lambda $ for the dynamical 
QCD scale for each case of different flavor numbers $N_{f}$, 
assuming that there will be no confusion.

Now we can put these curves into the Weierstrass form by changing the 
variables as $y=Y/2$ and $x=X + {\rm const. }$, i.e.,
\begin{eqnarray}
Y^2 =4(X-e_{1})(X-e_{2})(X-e_{3}),
\end{eqnarray}
where $e_{1}+e_{2}+e_{3}=0$.
According to Seiberg-Witten theory, the  coupling in (2) 
is given a very simple expression in terms of an integral over 
the two homology cycles, i.e.,
\begin{eqnarray}
\frac{\partial a(u,m_{i}, \Lambda)}{\partial u}&=&
\frac{\sqrt{2}}{4\pi}\oint _{\gamma _{1}}\frac{dX}{Y}=
\frac{\sqrt{2}}{2\pi}\frac{1}{\sqrt{e_{1}-e_{3}}}K(k),
\\
\frac{\partial a_{D}(u,m_{i}, \Lambda)}{\partial u}&=&
\frac{\sqrt{2}}{4\pi}\oint _{\gamma _{2}}\frac{dX}{Y}=
i\frac{\sqrt{2}}{2\pi}\frac{1}{\sqrt{e_{1}-e_{3}}}K(k').
\end{eqnarray}
Here the homology cycle $\gamma _{1}$ ($\gamma _{2}$) is defined 
as the one encircling 
the two points $e_{2}$ and $e_{3}$ ($e_{1}$). The function $K(k)$ 
is the complete elliptic integral of the first kind [11], i.e.,
\begin{eqnarray}
K(k)=\int _{0}^1 d\xi \frac{1}{\sqrt{(1-\xi ^2)(1-k^2 \xi ^2)}},
\end{eqnarray}
and we have also introduced 
\begin{eqnarray}
k^2=\frac{e_{2}-e_{3}}{e_{1}-e_{3}},
\end{eqnarray}
and 
$k'^2  = 1-k^2$.
In fact the  coupling in Eq. (2) is a simple function of 
a single variable $k^2$, i.e.,
\begin{eqnarray}
\tau \bigg (\frac{u}{\Lambda ^2}, \frac{m_{i}}{\Lambda}\bigg )=
\frac{\partial a_{D}(u, m_{i}, \Lambda )}{\partial u}\bigg /
\frac{\partial a(u, m_{i}, \Lambda )}{\partial u}
=i\frac{K(k')}{K(k)}.
\end{eqnarray}

The superconformal points are expected to exist 
 where the cycles shrink simultaneously. 
They are given by setting
\begin{eqnarray}
e_{1}=e_{2}=e_{3}=-(e_{1}+e_{2})=0.
\end{eqnarray} 
We will denote these points generically by $(u^{*}/\Lambda ^2, 
m_{i}^{*}/
\Lambda )$.
In order to argue  that the points $(u^{*}/\Lambda ^2, m_{i}^{*}/
\Lambda )$  are actually the superconformal points, an intuitive 
argument usually  goes in the following way. Suppose that we change 
the renormalization scale $M$ continuously and that 
$(u/\Lambda ^2, m_{i}/\Lambda )$ moves towards one of these points. 
If $u/\Lambda ^2$ and $m_{i}/\Lambda $ move in such a way that 
$k^2$ also goes towards some definite number. 
 Eq. (12) then shows that the coupling 
$\tau (u/\Lambda ^2, m_{i}/\Lambda )$ approaches some definite number 
independent of any mass scale. This means that the coupling ceases 
to move there and the theory becomes scale-invariant. 

This argument may be phrased in the following manner. Let us define 
our ``beta-function'' 
\begin{eqnarray}
\beta \bigg (\frac{u}{\Lambda ^2}, \frac{m_{i}}{\Lambda}\bigg )&=&
M\frac{d}{dM}\tau \bigg ( \frac{u}{\Lambda ^2}, \frac{m_{i}}{\Lambda} 
\bigg  )
\nonumber \\
&=&\bigg ( \gamma _{u}\frac{\partial}{\partial u}+ \sum _{i}\gamma _{mi}
\frac{\partial }{\partial m_{i}}\bigg )\tau  \bigg (\frac{u}{\Lambda ^2}, 
\frac{m_{i}}{\Lambda}\bigg ).
\end{eqnarray}
Here we have introduced two basic quantities
\begin{eqnarray}
\frac{\gamma _{u}}{\Lambda ^2}&=&M\frac{\partial }{\partial M}
\bigg (\frac{u}{\Lambda ^2}\bigg ),
\\
\frac{\gamma _{m_{i}}}{\Lambda}&=& M\frac{\partial }{\partial M}
\bigg (\frac{m_{i}}{\Lambda}\bigg ).
\end{eqnarray}
These are analogous to anomalous dimensions of 
$u/\Lambda ^2$ and $m_{i}/\Lambda $, respectively.

We are interested in the behavior of the beta-function near 
the superconformal points, and  we 
introduce a critical exponent of the beta-function or 
$\tau (u/\Lambda ^2 , m_{i}/\Lambda )$ in the infrared limit
\begin{eqnarray}
\beta \bigg (\frac{u}{\Lambda ^2}, \frac{m_{i}}{\Lambda}\bigg )
\sim {\rm const. } (M_{0}/M)^{-\rho },
\end{eqnarray}
or
\begin{eqnarray}
\rho = {\rm lim}_{M\rightarrow 0}\frac{\partial 
{\rm log}(\tau - \tau ^*)}{\partial {\rm log }M},
\end{eqnarray}
where we denoted $\tau (u^*/\Lambda ^2, m^* _{i}/\Lambda )$ by 
$\tau ^*$.
Since $\tau (u/\Lambda ^2, m_{i}/\Lambda )$ is a function of 
$k^2$ alone, Eq. (18) may be put into 
\begin{eqnarray}
\rho = {\rm lim }_{M\rightarrow 0}\frac{\partial {\rm log}
(k^2 - k^{*2})}{\partial {\rm log }M},
\end{eqnarray}
where $k^{*2}$ is the value of $k^2$ at the superconformal point.
We now study the exponent $\rho $ in the $SU(2)$ QCD with $N_{f}=1$, 
2, and 3 flavors and also in the $SU(3)$ pure Yang-Mills theory. 
\vskip1cm
\begin{flushleft}
{\bf 3. $SU(2), N_{f}=1$  case}
\end{flushleft}
\vskip0.5cm
In the simplest $N_{f}=1$ case, the solutions to $Y^2=0$ for the
elliptic curve given by (3), (4) and (7) turn out to be 
\begin{eqnarray}
e_{1}&=&\frac{1}{3}\bigg ( f^{1/3}_{+}+f^{1/3}_{-}\bigg ),
\\
e_{2}&=&\frac{1}{3}\bigg ( \omega f^{1/3}_{+} + \omega ^2 
f^{1/3}_{-}\bigg ),
\\
e_{3}&=&\frac{1}{3}\bigg ( \omega ^2 f^{1/3}_{+} + 
\omega f^{1/3}_{-} \bigg ).
\end{eqnarray}
Here $\omega ={\rm exp}(2\pi i/3)$ and we have introduced 
$f_{\pm }$ defined by 
\begin{eqnarray}
f_{\pm }&=&u^3 -\frac{9}{8}m\Lambda ^3 u +\frac{27}{128}\Lambda ^6 
\nonumber \\
& &\pm \frac{\sqrt{27}}{16}\Lambda ^3 \sqrt{4u^3-4m^2u^2-
\frac{9}{2}m\Lambda ^3 u+\frac{27}{64}\Lambda ^6 +4m^3 \Lambda ^3}.
\end{eqnarray}

The superconformal invariance is expected to exist at the
 points where mutually  non-local charges are present. 
They are obtained in this case   by imposing (13), 
or equivalently by taking the limit
 $f_{\pm }\rightarrow 0$.  More explicitly, they turn out to be 
\begin{eqnarray}
\bigg (\frac{u^{*}}{\Lambda ^2}, \frac{m^{*}}{\Lambda}\bigg )
=\bigg (\frac{3}{4},  \frac{3}{4} \bigg ),
\hskip0.5cm \bigg (\frac{3}{4}\omega, \frac{3}{4}\omega ^2 \bigg ),
 \hskip0.5cm \bigg (\frac{3}{4} \omega ^2, \frac{3}{4} \omega 
\bigg ).
\end{eqnarray}

As a matter of fact,  the function 
$\tau (u/\Lambda^2 , m/\Lambda )$ is ambiguous in general 
when $(u/\Lambda ^2 ,m/\Lambda)$ is going to one of these points.  
Eq. (11) shows that $k^2$ is in fact 0/0.  
The renormalization group flow should be such that the motion of
$(u/\Lambda ^2, m/\Lambda )$ under the change of $M$ conspires
 to give some definite numbers to $k^2$ and $\tau (u/\Lambda ^2,
 m/\Lambda )$. While the motion of $(u/\Lambda ^2, m/\Lambda )$
 is dictated by (15) and (16), their explicit 
forms are not known to us. The information of the two basic 
quantities, (15) and (16),  all comes from the high energy 
content  of the theory. 
This is also the case with the limits of $k^{2}$ and $\tau 
(u/\Lambda ^{2}, m/\Lambda )$.   
Their derivation starting from first 
principles is of course an  important problem.

In the following, however, we get around this problem by making 
use of the method of Argyres et al [4]. What they did amounts 
to the following. Let us parameterize the deformation from the 
superconformal points by setting
\begin{eqnarray}
\frac{m-m^{*}}{\Lambda}&=&D_{1}t^{-\alpha},
\\
\frac{u-u^{*}}{\Lambda ^2}&=&D_{1}^{\prime}t^{-\alpha }+D_{2}t^{-\beta}.
\end{eqnarray}
Here two numbers $\alpha $ and $\beta $ will be called scaling
 dimensions of $m$ and $u$, respectively. The flow 
$(u/\Lambda ^2, m/\Lambda )$ is parameterized by ``time'' $t$, 
and the limit $t\rightarrow +\infty $ drives the point 
$(u/\Lambda ^2, m/\Lambda )$ towards one of the superconformal 
points. Argyres et al.  have shown that  
one can adjust $k^{2}$ to any value, say, $k^{*2}$ by fine-tuning
if and only if we choose 
$\beta =3\alpha /2$ and $D_{1}=D_{1}^{\prime}$.

Let us now go one step further to identify the ``time ''
 variable with $M_{0}/M$.  The large time limit 
corresponds to the infrared limit 
$t=M_{0}/M \rightarrow +\infty $.
We mean by this that  the flow  of the moduli point 
$(u/\Lambda ^{2}, m/\Lambda )$ under the change of $t$ is the 
only possibility of the renormalization group flow.
This identification immediately enables us to derive the
formulae for (15) and (16), i.e.,
\begin{eqnarray}
\frac{\gamma _{m}}{\Lambda}&=&
\alpha \bigg (\frac{m-m^*}{\Lambda}\bigg ), 
\\
\frac{\gamma _{u}}{\Lambda ^2}&=&
\frac{3}{2} \alpha \bigg (\frac{u-u^*}{\Lambda ^2}
\bigg )- \frac{1}{2} 
\alpha \bigg (\frac{m-m^*}{\Lambda}\bigg ).
\end{eqnarray}
The beta-function can thus be evaluated by looking for 
various possible deformations around superconformal points.

With the ratio $\beta =3\alpha /2$, 
 Eq. (11) becomes the following form near the superconformal
 points
\begin{eqnarray}
k^2 - k^{*2} = {\rm const. }t^{-\alpha /2}.
\end{eqnarray}
Thus we can immediately conclude that the critical exponent 
$\rho $ becomes 
\begin{eqnarray}
\rho = \frac{1}{2}\alpha .
\end{eqnarray} 
If we normalize the 
critical exponent of $a(u, m, \Lambda )$ to be unity, 
the exponent $\alpha $ should be set equal to $4/5$ and 
thus we have $\rho  = 2/5.$  This result 
agrees with the corresponding quantity obtained by 
Bilal and Ferrari [9].

\vskip1cm
\begin{flushleft}
{\bf 4.  $SU(2), N_{f}=2$ Case}
\end{flushleft}
\vskip0.5cm
The calculational steps discussed in the $N_{f}=1$ case 
go through in the $N_{f}=2$ case as well with little 
modification. If we set the masses of hypermultiplets equal, 
$m_{1}=m_{2}$, we can expect higher criticality. In the 
following, we will keep our argument as general as possible 
and we introduce 
\begin{eqnarray}
m=\frac{1}{2}(m_{1}+m_{2}), \quad C_{2}=\frac{1}{2}(m_{1}-
m_{2})^{2},
\end{eqnarray}
instead of $m_{1}$ and $m_{2}$.

We have now three parameters, $u$, $m$ and $C_{2}$, while 
the condition for the superconformal points, 
Eq. (13), gives only two constraints.  
This indicates that the superconformal invariance 
is expected to exist along a 
line of complex-dimension one in the moduli space.
In order to scrutinize the one-dimensional line, let us
rewrite the elliptic curve in the form 
$Y^{2}=4(X^{3}+fX+g)$, where
\begin{eqnarray}
f&=&-\frac{1}{3}\left \{u^{2}-\left (\frac{3\Lambda ^{2}}{8}
\right )^{2}\right \}+\frac{\Lambda ^{2}}{4}\left \{m^{2}-\left (
\frac{\Lambda}{2}\right )^{2}\right \}-\frac{\Lambda ^{2}}
{8}C_{2},
\end{eqnarray}
\begin{eqnarray}
g&=&-\frac{2}{27}\left \{ u^{2}-\left (\frac{
3\Lambda ^{2}}{8}\right )^{2} \right \}u 
+ \frac{1}{12}\left (
u-\frac{3\Lambda ^{2}}{8}\right )m^{2}\Lambda ^{2}
\nonumber \\
& &-\left (\frac{1}{24}\Lambda ^{2}u + \frac{1}
{64}\Lambda ^{4}\right )C_{2}.
\end{eqnarray}

The point $(u^{*}/\Lambda ^{2}, m^{*}/\Lambda, 
C_{2}^{*}/\Lambda ^{2})$ on the line is the solution to the 
equations $f=0$ and $g=0$.  The analysis of the scaling 
dimensions goes exactly in the same way as in the 
$N_{f}=1$ case for generic values of 
$(u^{*}/\Lambda ^{2}, m^{*}/\Lambda , C_{2}^{*}/\Lambda ^{2})$
on  the line. Something peculiar happens, however, if we impose 
a further condition $C_{2}^{*}=0$, namely, at the points 
\begin{eqnarray}
\bigg (\frac{u^{*}}{\Lambda ^2}, \frac{m^{*}}{\Lambda}, 
\frac{C_{2}^{*}}{\Lambda ^{2}} \bigg )
=\bigg (\frac{3}{8},  \frac{1}{2}, 0 \bigg ),
\hskip0.5cm \bigg (\frac{3}{8},  -\frac{1}{2}, 0 \bigg ).
\end{eqnarray}
Here the two masses, $m_{1}$ and $m_{2}$, are set equal, and 
the first derivatives of the function $g$  with respect to 
$m$ and $u$ both vanish. The scaling behavior 
at this particular point differs from other points 
and we would like to concentrate ourselves  upon the point 
$(u^{*}/\Lambda ^{2}, m^{*}/\Lambda , C_{2}^{*}/\Lambda ^{2})=
(3/8, 1/2, 0)$ hereafter.

We use again the limiting procedure of Argyres et al. which
is in this case put in the following form
\begin{eqnarray}
\frac{m-m^*}{\Lambda}&=&D_{3}t^{-\alpha },
\\
\frac{u-u^*}{\Lambda ^2}&=&D_{3}^{\prime} t^{-\alpha}+D_{4}t^{-\beta}.
\end{eqnarray}
We should equally consider the deformation $C_{2}-C_{2}^{*}$, 
whose contributions are, however, always sub-leading and need 
not be  considered in the following discussions. 

We require again that $k^{2}$ can be  
 sent to a generic value $k^{*2}$ in the limit 
$t\rightarrow \infty $.
This is possible if and only if  $\beta =2\alpha $ and $D_{3}=
D_{3}^{\prime}$. As we did before identification $t=M_{0}/M$ gives us
immediately expressions of $\gamma _{u}$ and $\gamma _{m}.$
\begin{eqnarray}
\frac{\gamma _{m}}{\Lambda}&=&
\alpha \bigg (\frac{m-m^*}{\Lambda}\bigg ),
\\
\frac{\gamma _{u}}{\Lambda ^2}&=&
2\alpha \bigg (\frac{u-u^*}{\Lambda ^2}\bigg ) - 
\alpha \bigg (\frac{m-m^*}{\Lambda }\bigg ).
\end{eqnarray}

Putting (35) and (36) into Eq. (11), the infrared 
behavior of $k^2$  is approximated as
\begin{eqnarray}
k^2 = \bigg (\frac{1}{2}-\frac{D_{3}}{\sqrt{D_{4}}}
\bigg ) + \frac{D_{3}^2  - D_{4}}{2\sqrt{D_{4}}}t^{-\alpha }
+ \cdots \cdots.
\end{eqnarray}
This shows us the relation of scaling dimensions 
\begin{eqnarray}
\rho = \alpha .
\end{eqnarray} 
The normalization of the scaling dimension of 
$a(u, m_{i}, \Lambda )$ being equal to unity leads us to the 
relation  $\rho = 2/3$, which  agrees again with the 
results obtained by Bilal and Ferrari [9].

\vskip1cm
\begin{flushleft}
{\bf 5.  $SU(2)$, $N_{f}=3$ Case}
\end{flushleft}
\vskip0.5cm
Let us turn to the $N_{f}=3$ case.  We consider the 
combination of the bare  masses of matter hypermultiplets
\begin {eqnarray}
m=\frac{1}{3}\sum _{i=1}^{3} m_{i}, \quad 
C_{2}=\sum _{i=1}^{3} (m_{i}-m)^{2}, \quad  
C_{3}=\sum _{i=1}^{3} (m_{i}-m)^{3},
\end{eqnarray}
instead of $m_{1}$, $m_{2}$ and $m_{3}$. 
In terms of these, the elliptic curve given by (6) and 
(7) is put into the form
$Y^{2}=4(X^{3}+{\tilde f}X+{\tilde g}),$
where
\begin{eqnarray}
{\tilde f}&=&-\frac{1}{3}\left \{ u-2\left (\frac{\Lambda}{8}
\right )^{2}\right \}^{2}
+ \left (\frac{\Lambda}{8}\right )\left (
m-\frac{\Lambda}{8}\right )^{2}\left ( 2m+\frac{\Lambda}
{8}\right )
\nonumber \\
& & -C_{2}\left \{m\left (\frac{\Lambda }{8}\right )+
\left (\frac{\Lambda}{8}\right )^{2}\right \}
+\frac{2}{3}\left (\frac{\Lambda}{8}\right )C_{3},
\end{eqnarray}
\begin{eqnarray}
{\tilde g}&=&-\frac{2}{27}\left \{u-2\left (\frac{\Lambda}{8}
\right )^{2}\right \}^{2}\left \{ u+\frac{11}{8}\left (
\frac{\Lambda}{8}\right )^{2}\right \}
-3\left (\frac{\Lambda}{8}\right )^{2}\left (
m^{2}-\frac{u}{2} \right )^{2}
\nonumber \\
& &+\frac{1}{3}\left \{ u+\left ( \frac{\Lambda}{8}
\right )^{2}\right \}\Bigg [
2\left (\frac{\Lambda}{8}\right )\left ( m-\frac{
\Lambda}{8}\right )^{2}\left \{ m+\frac{1}{2}\left (
\frac{\Lambda}{8}\right )\right \}
\nonumber \\
& &\qquad \qquad \qquad +\frac{2}{3}\left (
\frac{\Lambda}{8}\right ) C_{3}
-C_{2}\left \{ m\left (\frac{\Lambda}{8}\right )
+\left (\frac{\Lambda}{8}\right )^{2}\right \}
\Bigg ]
\nonumber \\
& &+\left (\frac{\Lambda}{8}\right )^{2}uC_{2}
+\left (\frac{\Lambda}{8}\right )^{2}\left (
2mC_{3}-\frac{C_{2}^{2}}{4}\right ).
\end{eqnarray}

The condition (13) for the superconformal invariance 
to exist is given by the solutions to the equations 
${\tilde f}=0$ and ${\tilde g}=0$. These equations define    
a surface of complex-dimension two in the moduli space. 
The analysis of the scaling dimensions goes in a similar 
way on the generic point on the surface, as was done 
in $N_{f}=1$ and $N_{f}=2$ cases.  As we realize  rather 
easily from the expression of (42) and (43), however, 
a different type of scaling is expected at
\begin{eqnarray}
\left (
\frac{u^{*}}{\Lambda ^{2}}, \frac{m^{*}}{\Lambda}, 
\frac{C_{2}^{*}}{\Lambda ^{2}}, \frac{C_{3}^{*}}
{\Lambda ^{3}}
\right )=\left (
\frac{1}{32}, \frac{1}{8}, 0, 0
\right ).
\end{eqnarray}
Actually, the first derivatives of ${\tilde f}$ and 
 ${\tilde g}$ with respect to $u$ and $m$ all vanish, 
and the critical behavior becomes different. 

Let us restrict ourselves to this particular point (44) 
and  we again parameterize the deformation around (44), i.e.,

\begin{eqnarray}
\frac{m-m^{*}}{\Lambda }&=&D_{5}t^{-\alpha},
\\
\frac{u-u^{*}}{\Lambda ^{2}}&=&D_{5}^{\prime}t^{-\alpha}+
D_{5}^{\prime \prime}t^{-\beta}+D_{6}t^{-\gamma}.
\end{eqnarray}
The deviations $C_{2}-C_{2}^{*}$ and $C_{3} -
C_{3}^{*}$ should also be given due consideration, but 
they  will turn out to be  irrelevant to our following 
consideration.

Suppose that we would like to be able to tune 
the value  $k^2$ to a generic  number $k^{*2}$. This  
is possible if and only if $D_{5}^{\prime}=3D_{5}/
8$, $D_{5}^{\prime \prime}=D_{5}^{2}$, $\beta =2\alpha$, 
and $\gamma = 3\alpha$.
As a record,  we write down the functional form of 
$\gamma _{u}$ and $\gamma _{m}$ in this case;
\begin{eqnarray}
\frac{\gamma _{m}}{\Lambda}&=&
\alpha \bigg (\frac{m-m^*}{\Lambda}\bigg ),
\end{eqnarray}
\begin{eqnarray}
\frac{\gamma _{u}}{\Lambda ^2}&=&
3\alpha \bigg (\frac{u-u^*}{\Lambda ^2}\bigg )-\frac{3}{4}
\alpha \bigg (\frac{m-m^*}{\Lambda}\bigg )
- \alpha \bigg (\frac{m-m^*}{\Lambda }\bigg )^2.
\end{eqnarray}

As we did before, we put (45) and (46) into Eqs. (20)-(23) 
 and we find the leading 
behavior of $k^2$ near its value $k^{*2}$ at superconformal points
\begin{eqnarray}
k^2 - k^{*2} = {\rm const. }t ^{-\alpha }.
\end{eqnarray}
This shows us  the relation
\begin{eqnarray}
\rho = \alpha .
\end{eqnarray}
The scaling dimension of $a(u, m_{i}, \Lambda )$, $2\alpha $ 
being set equal to unity, we get $\rho =1/2$.
This, however, differs from the corresponding result of 
Bilal and Ferrari [9] by factor 2. 
We are not quite sure of the origin
of this discrepancy.

\vskip1cm
\begin{flushleft}
{\bf 6.  The $SU(3)$  Yang-Mills Theory}
\end{flushleft}
\vskip0.5cm
Finally let us come to the case of $SU(3)$ gauge group broken to 
$U(1)\times U(1)$  in the 
$N=2$ supersymmetric pure Yang-Mills theory.
In this case we begin with the hyper-elliptic curve [12-14]
with two moduli parameters, $u$ and $v$, 
\begin{equation}
   y^2 = (x^3 - u x -v)^2 - \Lambda^6 =\prod _{i=1}^{6}(x-e_{i}).  
\end{equation}
The branch points $e_{i}$  are given by
\begin{eqnarray}
   e_{1} = f_{1 +} + f_{1 -} ,\! \qquad  &\quad& 
   e_{4} = f_{2 +} + f_{2 -},\\
   e_{2} = \omega f_{1 +} + \omega^2 f_{1 -} , &\quad& 
   e_{5} = \omega f_{2 +} + \omega^2 f_{2 -},    \\
   e_{3} = \omega^2 f_{1 +} + \omega f_{1 -} , &\quad& 
   e_{6} = \omega^2 f_{2 +} + \omega f_{2 -},  
\end{eqnarray}
where $\omega ={\rm exp}(2\pi i /3)$ and 
our other notations are
\begin{eqnarray} 
   f_{1 \pm} &=& 2^{-1/3} \left \{ (v - \Lambda^3) \pm 
   \sqrt{(v-\Lambda^3)^2-\frac{4}{27} u^3} \right \}^{1/3},   
\\
   f_{2 \pm} &=& 2^{-1/3} \left \{ (v + \Lambda^3) \pm 
   \sqrt{(v+\Lambda^3)^2-\frac{4}{27} u^3} \right \}^{1/3}. 
\end{eqnarray}

The low-energy  coupling $\tau ^{ij}$ is given in terms of  
 $2\times 2$  matrices 
\begin{eqnarray}
   \tau ^{ij} = {\bf B}\cdot{\bf A}^{-1}, 
\end{eqnarray}
where
\begin{eqnarray}
   \qquad  {\bf A}= \left(
      \begin{array}{lr}
         \partial_{v} a_{1}  &  \partial_{u} a_{1} \\
         \partial_{v} a_{2} & \partial_{u} a_{2}
      \end{array} \right) &, &
   {\bf B}= \left(
      \begin{array}{lr}
         \partial_{v} a_{D}^{1}  &  \partial_{u} a_{D}^{1} \\
         \partial_{v} a_{D}^{2} & \partial_{u} a_{D}^{2}
      \end{array} \right).
\end{eqnarray}
One can get the explicit form of each  matrix element of 
${\bf A}$ and ${\bf B}$  from the exact solution, to find
\begin{eqnarray}
\frac{\partial a_{i}}{\partial u} = \frac{\sqrt{2}}{4\pi}
\oint_{\alpha_{i}}\frac{x dx}{y} ,
&\quad&
\frac{\partial a_{i}}{\partial v} = \frac{\sqrt{2}}{4\pi}
\oint_{\alpha_{i}}\frac{dx}{y}, 
 \\
\frac{\partial a_{D}^{i}}{\partial u} = \frac{\sqrt{2}}{4\pi}
\oint_{\beta_{i}}\frac{xdx}{y},
&\quad &
\frac{\partial a_{D}^{i}}{\partial v} = 
\frac{\sqrt{2}}{4\pi}\oint_{\beta_{i}}\frac{dx}{y}. 
\end{eqnarray}
We will use the same definition of the cycles $\alpha _{i}$ and 
$\beta _{i}$ as was used by Argyres and Douglas [3]: 
The $\alpha _{1}$  ($\beta _{1}$) cycle is the one encircling 
 $e_{3}$ and $e_{2}$ ($e_{1}$), and the $\alpha _{2}$ 
($\beta _{2}$) cycle the one encircling 
$e_{6}$ and $e_{5}$ ($e_{4}$).

Let us think of a special situation in which massless particles with 
mutually non-local charge appear. This situation corresponds to the 
existence of superconformal symmetry and is realized if 
 $a_{1} = a_{D}^{1} = 0$, or  $a_{2} = a_{D}^{2} = 0$.
In other words,  two points  ($u^{*}/\Lambda ^{2}$, 
$v^{*}/\Lambda ^{3}$)=($0$, $\pm 1$) are nothing but 
the superconformal points. Without losing generality, we confine 
ourselves  to only  one of the two, namely,  
($u^{*}/\Lambda ^{2}$, $v^{*}/\Lambda ^{3}$)=($0$, $1$).
The branch points, Eqs.  (52)-(54),  for 
this particular choice of $u$ and $v$ turn out to be 
\begin{eqnarray}
e_{1}^{*} = e_{2}^{*} = e_{3}^{*}= 0,
     \quad 
(e_{4}^{*}, e_{5}^{*}, e_{6}^{*}) =
 2^{1/3}\Lambda (1, \omega ,  \omega^2 ) .
\end{eqnarray} 
We would like to analyze the behavior of $\tau ^{ij}$ 
in the vicinity of  this superconformal point.
By changing integration  variables in (59) and (60), 
we get a more useful 
form  amenable to  further analysis.
\begin{eqnarray}
\frac{\partial a_{1}}{\partial v}& = &\frac{\sqrt{2}}{4\pi}
\frac{4}{\sqrt{(e_{2}-e_{1})(e_{4}-e_{2})(e_{5}-e_{2})(e_{6}-e_{2})}}
\nonumber \\
& & \times  \int_{0}^{1} d\xi  \frac{1}{\sqrt{(
1-\xi ^2)(1-k^2 \xi ^2)(1- \delta_{4} \xi ^2)(1- \delta_{5} 
\xi ^2 )(1- \delta_{6} \xi ^2 )}},
\\ 
\frac{\partial a_{1}}{\partial u}& = &\frac{\sqrt{2}}{4\pi}
\frac{4}{\sqrt{(e_{2}-e_{1})(e_{4}-e_{2})(e_{5}-e_{2})(e_{6}-e_{2})}} 
\nonumber \\ 
& & \times  \int_{0}^{1} d\xi  \frac{(e_{3}-e_{2})\xi ^2 + e_{2}}
{\sqrt{(1-\xi ^2)(1-k^2 \xi ^2)(1- \delta_{4} \xi^2 )
(1- \delta_{5} \xi ^2 )(1- \delta_{6} \xi ^2 )}}, 
\end{eqnarray}
where $k^2 = (e_{3} - e_{2})/(e_{1} - e_{2})$ and  
$\delta_{i} = (e_{3} - e_{2})/(e_{i} - e_{2}),  \quad (i = 4,5,6)$.
We obtain $\partial_{v} a_{D}^{1}$, and $\partial_{u} a_{D}^{1}$  
in a similar way  by  exchanging $e_{1}$ and $e_{2}$. In other words,
  $k^2$ and $\delta _{i}$ in (62) and (63) should be replaced by  
$k^{\prime 2} = 1 - k^2$ , and  $\delta^{\prime}_{i} =(e_{3}-
e_{1})/(e_{i} - e_{1})$, respectively.

The above concise expression enables us to  evaluate the infrared
behavior of (59) and (60) near the superconformal point by setting 
$k^2 \rightarrow k^{* 2}$, $k^{\prime 2}\rightarrow k^{\prime * 2}$, 
$\delta _{i} \rightarrow 0$, and
$\delta ^{\prime}_{i} \rightarrow 0$, i.e.,
\begin{eqnarray}
\frac{\partial a_{1}}{\partial v} \longrightarrow 
\frac{1}{\pi \Lambda ^{3/2}}\frac{1}
{\sqrt{e_{2}^{*}-e_{1}^{*}}} K(k), 
&\quad&
\frac{\partial a_{D}^{1}}{\partial v} \longrightarrow  
\frac{1}{\pi \Lambda ^{3/2}}\frac{i}
{\sqrt{e_{2}^{*}-e_{1}^{*}}}K(k^{\prime }),
\\
\frac{\partial a_{1}}{\partial u}\longrightarrow 0, 
&\quad &
\frac{\partial a_{D}^{1}}{\partial u}\longrightarrow 0.
\end{eqnarray}
Eq. (64)  is reminiscent of the $SU(2)$ case, Eqs. (8) 
and (9).
Let us now have a closer look at the limiting behavior (64) and (65)
near the superconformal point, following the  argument of the work
of Argyres et al [4]. We consider the deviation from the 
superconformal point by introducing a parameter $t$, i.e.,
\begin{equation}
   \frac{u- u^{*}}{\Lambda ^{2}} = E_{1} t^{-\alpha}, 
\quad 
   \frac{v- v^{*}}{\Lambda ^{3}} = E_{2} t^{-\beta }. 
\end{equation} 
The ratio of the exponents $\beta /\alpha $ should be determined 
in order that  the coupling $\tau ^{11}$ should go to a 
generic  value.  
This condition provides us uniquely with  $\beta = 3\alpha /2$.  
Plugging this result, one can evaluate the asymptotic 
behavior of each  integral, i.e.,
\begin{eqnarray}
\frac{\partial a_{1}}{\partial v} &\sim & \frac{1}{\pi \Lambda ^{3/2}}
\frac{1 }{\sqrt{e_{2}^{*}-e_{1}^{*}}}\left \{ K(k) + 
{\cal O}(t^{-3 \alpha /2}) \right \},
\\
\frac{\partial a_{D}^{1}}{\partial v} &\sim & \frac{1}{\pi \Lambda ^{3/2}}
\frac{ i}{\sqrt{e_{2}^{*}-e_{1}^{*}}}\left \{ K(k^{\prime }) + 
{\cal O}(t^{-3 \alpha /2})\right \},
\\
\frac{\partial a_{1}}{\partial u} &\sim & {\cal O}(t^{- \alpha /4}),
\\
\frac{\partial a_{D}^{1}}{\partial u} &\sim & {\cal O}
(t^{- \alpha /4}).
\end{eqnarray}

In the same way, the other integrals associated with the bigger 
homology cycle turn out to be
\begin{eqnarray}
\frac{\partial a_{2}}{\partial v} &=&\frac{\sqrt{2}}{4\pi}
\frac{4}{\sqrt{(e_{4}^{*}-e_{5}^{*}) 
(e_{6}^{*}-e_{5}^{*})^{3}}} \left \{I_{1} + {\cal O}(t^{-\alpha})\right \} , 
\\
\frac{\partial a_{D}^{2}}{\partial v} &=& \frac{\sqrt{2}}{4\pi}
\frac{4i}{\sqrt{(e_{4}^{*}-e_{5}^{*})
(e_{6}^{*}-e_{4}^{*})^{3}}} \left \{ I_{2} + {\cal O}(t^{-\alpha})\right \}, 
\\
\frac{\partial a_{2}}{\partial u} &=& \frac{\sqrt{2}}{4\pi}
\frac{4}{\sqrt{(e_{4}^{*}-e_{5}^{*})
(e_{6}^{*}-e_{5}^{*})}} \left \{J_{1} + {\cal O}(t^{-\alpha})\right \}, 
\\
\frac{\partial a_{D}^{2}}{\partial u} &=& \frac{\sqrt{2}}{4\pi}
\frac{4 i}{\sqrt{(e_{4}^{*}-e_{5}^{*})
(e_{6}^{*}-e_{4}^{*})}}\left \{J_{2} + {\cal O}(t^{-\alpha})\right \}, 
\end{eqnarray}

where 
\begin{eqnarray}
I_{1} &=& \int_{0}^{1} d\xi \frac{1}{\sqrt{(1-\xi ^2)(1- \ell ^{2} 
\xi ^2) (\xi ^2 + \kappa)^{3}}}, 
 \\
J_{1} &=& \int_{0}^{1}  d\xi \frac{1}{\sqrt{(1-\xi^2)(1- \ell ^{2}
\xi ^{2})(\xi^2 + \kappa)}}. 
\end{eqnarray}
Here our notations are $\ell ^{2}=-\omega $, and   $\kappa = 1/(
\omega - 1)$.
One can also obtain the expressions of $I_{2}$ and $J_{2}$ by 
replacing $\ell ^2$ and $\kappa $ in (75) and (76) by $\ell ^{\prime 2} = 
1-\ell ^2 =-\omega ^{2}$  and $\kappa ^{\prime }=1/(\omega^2 - 1)$, 
respectively.  
Simple relations  $I_{2}=I_{1}^{*}$ and $J_{2}=J_{1}^{*}$ 
immediately come to our notice, because of 
$\ell ^{\prime } =\ell ^{*}$ and $\kappa ^{\prime }=\kappa ^{*}$.
It is also amusing to see that the evaluation $J_{2}/J_{1}=J_{1}^{*}
/J_{1}$  is analytically possible, i.e.,  
$J_{2}/J_{1}=(1 - \sqrt{3}i)/2$. 

By using the above asymptotic form of the  integrals (67)-(74),
we have analyzed the asymptotic behavior of the effective coupling 
$\tau ^{ij}$. Our final result is as shown below:
\begin{eqnarray}
   \tau ^{11} &=& i \frac{K(k^{\prime })}{K(k)} + 
   {\cal O}(t^{- \alpha /2}) , 
\\
   \tau ^{22} &=& i e^{i \pi /6} \left(\frac{J_{2}}
   {J_{1}}\right) +{\cal O}(t^{- \alpha /2})
\nonumber \\
   &=&e^{i \pi /3}  + {\cal O}(t^{- \alpha /2}) ,
\\
   \tau ^{12} &=& \tau ^{21} =  {\cal O}(t^{- \alpha /4}).
\end{eqnarray} 
We are thus led  to conclude that the slope parameters of 
the beta-function
associated with the coupling $\tau ^{ij}$ are 
\begin{eqnarray}
\rho = \alpha /2 &{\rm for}& \tau ^{11}\quad {\rm and } 
\quad \tau ^{22},
\\
\rho =\alpha /4 &{\rm for}& \tau ^{12} \quad {\rm and } 
\quad \tau ^{21}.
\end{eqnarray}
The exponent of $a_{1}$ being unity as in the $SU(2)$, 
$N_{f}=1$ case, $\alpha $ should be set equal to 4/5. This leads
 us to $\rho =2/5$ for $\tau ^{11}$ and $\tau ^{22}$, and $\rho =
1/5$ for $\tau ^{12}$ and $\tau ^{21}$.

The difference between  $\tau ^{11}$ and $\tau ^{22}$ is
that, while the infrared limit of 
$\tau ^{22}$ has been determined by (78) uniquely, 
 the value of $\tau ^{11}$ at the superconformal point 
 is not :  the value of $\tau  ^{11}$ there depends on the coefficients
$E_{1}$ and $E_{2}$ in (66), whose determination belongs to procedure 
of integrating out the high-energy content of the theory.
These  facts could be connected with the observation  that $U(1)$ 
gauge theory with coupling  $\tau ^{11}$ is an interacting 
non-trivial superconformal field theory, while $U(1)$ gauge 
theory with coupling $\tau ^{22}$ is a free trivial one.  

Argyres et al. claim that the superconformal field theory 
considered in the present Section i.e.,  with $SU(3)$ 
pure Yang-Mills theory 
is the same  (1, 1) type as that of  the 
superconformal field theory  of  $SU(2)$ QCD with one flavor.
Our result involving the scaling dimension of $\tau ^{11}$ 
is a non-trivial check of their argument.

\vskip1cm
\begin{flushleft}
{\bf 7.  Summary and Discussions }
\end{flushleft}
\vskip0.5cm
In the present paper we have investigated the renormalization group 
flow near the superconformal points in $SU(2)$ QCD and $SU(3)$ pure 
Yang-Mills theory. We have determined the slope parameter (18) of the 
beta-function, which agrees partly with results obtained previously by 
Bilal and Ferrari in a different method.

As we mentioned briefly in Introduction, there have been  a lot of 
elaborate works on the renormalization group equation in the 
Seiberg-Witten theory [6, 8, 15]. It has been often 
argued in literatures that taking the derivative 
of $\tau $ with respect to the QCD dynamical scale $\Lambda $ 
reproduces the correct beta-function in the weak coupling region.
 Since the coupling $\tau $ in the $SU(2)$ case 
is a function of  $u/\Lambda ^{2}$ and $m_{i}/\Lambda $,
the beta-function in the weak coupling region turns out to be 
expressed by
\begin{eqnarray}
-\Lambda \frac{\partial }{\partial \Lambda }\tau \left (\frac{u}
{\Lambda ^{2}}, \frac{m_{i}}{\Lambda}\right )=
\left \{ 2u\frac{\partial}{\partial u}+\sum _{i} m_{i}\frac{\partial }
{\partial m_{i}}\right \}\tau  \left (\frac{u}
{\Lambda ^{2}}, \frac{m_{i}}{\Lambda}\right ).
\end{eqnarray}
This indicates that our two basic quantities $\gamma _{u}$ and 
$\gamma _{m i}$ used extensively in our work  are given in the 
semi-classical region by $2u $ and $m_{i}$, respectively.
In the strong coupling region, on the other hand, we have successfully
 determined the functional form of $\gamma _{u}$ and $\gamma _{mi}$ 
from the consistency argument.  We have, however, no convincing way 
to  obtain $\gamma _{u}$ and $\gamma _{mi}$ in the whole region 
of the moduli space. The procedure of integrating out the high energy 
content of the theory above $M$ have hidden the information 
of the full $M$ dependence of $u$ and $m_{i}$.  The point
 of our work is that, despite the lack of the full information of the 
$M$-dependence, we are still able to determine  the renormalization 
group flow near the superconformal point by looking for possible
 deformation thereof.
\vskip1.5cm

\begin{flushleft}
{\bf Acknowledgments} 
\end{flushleft}

The authors would like to express their  sincere thanks 
to Luis {\' A}lvarez-Gaum{\' e} for numerous helpful discussions.
 One of the authors (T.K.) would like to thank Drs. G. Veneziano 
and A. De R{\' u}jula for their warm hospitality during his stay at 
CERN Theory Division in summer 1997, where part of this work 
was done.   This work was supported in part by Scientific Grants 
from the Ministry of Education (Grant Number, 09640353)
\vskip1cm
\begin{flushleft}
{\bf References}
\end{flushleft}
\begin{description}
\item{[1]}
N. Seiberg and E. Witten, Nucl. Phys. {\bf B426} (1994) 19; 
  {\bf B 430} (1994) 485 (E);  {\bf B431} (1994) 484.
\item{[2]}
W. Lerche, Nucl. Phys. B (Proc. Suppl. ) {\bf 55B} (1997) 83 
(hep-th/9611190);
L. {\' A}lvarez-Gaum{\' e} and S.F.  Hassan, 
Fortsch. Phys. {\bf 45}  (1997)  159  ( hep-th/9701069); 
L. {\' A}lvarez-Gaum{\' e} and F. Zamora, hep-th/9709180;
A. Bilal, hep-th/9601007;
C. G{\' o}mez and R. Hernandez, hep-th/9510023.
\item{[3]}
P. Argyres and M. Douglas, Nucl. Phys. {\bf B448} (1995) 93.
\item{[4]}
P. Argyres, M. Plesser, N. Seiberg and E. Witten, Nucl. Phys. 
{\bf B461} (1996) 71.
\item{[5]}
T. Eguchi, K. Hori, K. Ito  and S.-K. Yang, Nucl. Phys. {\bf B471} 
(1996) 430; 
T. Eguchi and K. Hori, in  {\it ``Mathematical 
Beauty of Physics:  A Memorial Volume for Claude Itzykson''} 
(ed. by J.M. Drouffe and J.B. Zuber, World Scientific Pub., 1997)
 p. 67.
\item{[6]}
G. Bonelli and M. Matone, Phys. Rev. Lett. {\bf 76} (1996) 4107.
\item{[7]}
J. Sonnenschein, S. Theisen and S. Yankielowicz, Phys. Lett. 
{\bf B367} (1996) 145; 
\item{[8]}
J.A. Minahan and D. Nemeschansky, Nucl. Phys. {\bf B 468} (1996)
72. 
\item{[9]}
A. Bilal and F. Ferrari, hep-th/9706145; 
F. Ferrari and A. Bilal, Nucl. Phys. {\bf B 469} (1996) 387;
A. Bilal and F. Ferrari, Nucl. Phys. {\bf B 480} (1996) 589. 
\item{[10]}
M. Matone, Phys. Lett. {\bf B 357} (1995) 342; Phys. Rev. 
{\bf D 53} (1996) 7354;
T. Eguchi and S.-K. Yang, Mod. Phys. Lett.  {\bf A 11} (1996) 131;
E. D'Hoker, I.M. Krichever and D.H. Phong, Nucl. Phys. {\bf B 494} 
(1997) 89.
L. {\' A}lvarez-Gaum{\' e}, M. Marino and F. Zamora, hep-th/9703072, 
hep-th/9707017.
H. Itoyama and A. Morozov, Nucl. Phys. {\bf  B 477} (1996) 855;
{\bf B 488 } (1996) 1.
T. Nakatsu and K. Takasaki, Mod. Phys. Lett. {\bf A 11} (1996) 157.
\item{[11]}
A. Hurwitz und R. Courant, {\it Vorlesungen {\"u}ber allgemeine 
Funktionentheorie und elliptische Funktionen}  (Springer-Verlag, 
Berlin, 1925).
\item{[12]}
P. Argyres and A.E. Faraggi, Phys. Rev. Lett. {\bf 74} 
(1995) 3931.
\item{[13]}
A. Klemm, W. Lerche, S. Theisen and  S. Yankielowicz, 
Phys. Lett. {\bf B 344} (1995) 169;
A. Klemm W. Lerche and  S. Theisen, Int. J. Mod. Phys. {\bf A11} 
(1996) 1929.
\item{[14]}
A. Hanany and Y. Oz, Nucl. Phys. {\bf B 452} (1995) 283.
\item{[15]}
A. Ritz, hep-th/9710112; B.P. Dolan, hep-th/9710161; 
J.I. Latorre and C.A. L{\" u}tken, hep-th/9711150;
G. Bonelli and M. Matone, hep-th/9712025.
\end{description}
\end{document}